# Stocks recommendation strategy based on a comparison between large number of stocks

Kartikay Gupta[a]* and Niladri Chatterjee[a]

*Abstract*

Stock return forecasting is of utmost importance in the business world. This has been the favourite topic of research for many academicians for decades. Recently, regularization techniques have reported increasing the forecast accuracy of the simple regression model tremendously. Still, these models cannot incorporate the effect of unpredictable things like a major natural disaster, large foreign influence, etc. in their prediction. They simply try to predict future values based on past values of each time series. Thus, it is more important to recommend top stocks rather than predicting exact stock returns. The present paper modifies the regression task to output value for each stock, which is more suitable for ranking the stocks by expected returns. The ranking is done out of the comparison between the stocks in the previous quarter. Two large datasets consisting of altogether 1205 companies listed at Indian exchanges were used for experimentation. Stochastic Gradient Descent (SGD) technique is used in this work to train the parameters, which allows scalability to even larger datasets. Five different metrics were used for evaluating the different models. Results were also analysed subjectively through plots. The results showed the superiority of the proposed techniques.

*Keywords: equity returns, forecasting, fundamentals, regression, ranking*

## 1. Introduction

Forecasting of equity price is very important for portfolio management and investment purposes. By rough estimates , around 6,30,000 companies are publicly traded around the world. Over the last decade or so a large number of techniques have been proposed to deal with the problem of forecasting company's share prices or equity over different horizons. Some of these are [1]–[7]. These researchers have used various indicators, including technical and fundamental ones, for forecasting. The consensus that may be derived from these studies is that equity premium is predictable over different horizons. The present study corroborates the above conclusion. It further proposes a novel technique which uses Kitchen-sink regression [8] and a big set of fundamental, technical and general indicators. 'Kitchen-sink' regression puts all the possible independent variables into the regression equation to be able to predict the values of the dependent variable. The regression parameters are trained in such a way that model score should rank the companies in the order of the next quarter returns. In this approach, no importance is given to predicting the actual return while training the model. But it is expected that a company with a higher model score should give higher next quarter return. This technique should nullify the effect of unpredictable factors such as weather conditions or foreign influence on the national economy. This proposed technique will be referred to as rank-regression (RR) as here the regression-output, is designed to rank the samples within a given list, rather than predicting each samples target values. This techniques will, be further elaborated in the next section. As explained through experimentation, the above strategy recommends top companies which give better performance than other prevalent techniques. The top companies recommended through the proposed models gave higher quarterly returns than other models when compared over large datasets.

The rest of the paper is organised as follows. Section 2 provides a brief literature review of the subject. Section 3 explains the proposed rank-regression technique and evaluation metrics. Section 4 gives the results of the experiments. Section 5 discusses the finer points of experimental results which demonstrate the superiority of proposed models. Section 6 presents the conclusion.

## 2. Literature Review

The present paper utilises a technique built over Kitchen-sink regression to forecast equity returns conditioned on a big set of important variables. Kitchen-sink regression uses all available explanatory variables to form a linear function which tries to output target values. Suppose $(X^i, y^i)$, i.e., [1, 2, .......N] are data points, on which kitchen

[a]*Mathematics Department, IIT Delhi, New Delhi, India*

*\*corresponding author details: Kartikay Gupta, Mathematics Department, IIT Delhi, New Delhi, Pin:110016, India*

e-mail: maz158144@iitd.ac.in.

sink regression needs to be done. Here $X^i = (x_1^i, x_2^i, \ldots, x_n^i)$ is a vector of explanatory variables and y is the corresponding target. Then, it tries to solve the following linear equations, simultaneously.

$$\sum_{j=1}^{n} p_j \times x_j^i = y^i, \quad for\ each\ \ i \in \{1, 2, \ldots N\}$$

In general, the exact solution may not exist, but the parameters ($p_i$), are obtained such that the following loss function is minimized.

$$\text{Loss} = \sum_{j=1}^{n}\left(y^i - \sum_{j=1}^{n} p_j x_j^i\right)^2$$

The function $\sum_{j=1}^{n} p_j x_j^i$ can then be used to estimate the unknown $y^i$.

Various research papers are available [9]–[11], which analyse the impact of some subset of the used variables on equity returns. In the study conducted in [11], it was empirically shown that there exists a positive relation between stock return and the ratio of debt to equity. Statistically significant relationship between equity return and four company fundamentals variables (market capitalization, book equity to market equity ratio, price earnings ratio and debt-equity ratio) was shown in [10]. Bhar et al [9] empirically show existence of statistically significant relation between oil price returns and Indian equity returns. This relationship is often negative. Since India imports its oil, higher oil prices leads to higher import bills of Indian companies, which eventually leads to lower stock returns. They also suggested that Indian economy generally remains unaffected by outside global events.

In work done in [2] the performance of certain important variables in equity premium prediction was analyzed. They found that those important variables alone are not robust for profitable and timely prediction. We agree on the point that it is undoubtedly difficult to obtain good forecasts, but some variables do contain useful information as would be discussed in this paper. Their findings were contoured in the research work [3] and [1]. It was shown empirically in [3] that many predictive variables beat the historical average return once weak restrictions are put on the sign of coefficients and variables. Many variables were combined in [1] to obtain forecasts using the combined power of individual variables. They empirically showed that their strategy was useful in reducing forecast volatility while incorporating information from several variables in the forecasts. The present study also uses the combined power of several variables to make equity forecasts.

The existence of out of sample predictability in equity premium risk was shown in [12]. They used a stochastic linear regression model to forecast monthly returns of the S&P 500 index. They modelled the coefficients as the random walk model and allowed them to vary over time. The present study does not allow that kind of uncertainty in the model and rather keeps any uncertainty to be included only in the error term. This keeps the model simple and precise while still showcasing its effectiveness. In the study conducted in [5], the importance of combining several technical and economic fundamental variable in obtaining improved forecasts for equity return was shown.

Jiahan et al. [13] also used Kitchen-sink regression to forecast exchange rates conditioned on economic fundamentals. A recent work [6] also used Kitchen sink regression to forecast equity returns. They took the forecast horizon of one month and used monthly economic fundamental indicators. While training the parameters for the regression model they made the sign of the coefficient same as the theoretically motivated sign between the equity fundamentals and the expected return. This was done by first obtaining the coefficients, and then those coefficients were made equal to zero, which did not have the desired sign. Secondly, they added L1 and L2 regularisation term to the final error, so that the coefficients remain small. In L1 regularisation, the absolute value of the coefficients is also minimised along with the square-loss function. In L2 regularisation, the squared value of the coefficients is minimised along with the loss function. Their findings were that equity premium is predictable out-of-sample.

## 3. Methodology

We have used a set of 25 indicators (as explained in Table 3A, Appendix) for forecasting one quarter ahead relative-return for each company. The quarterly-return value used to generate relative-return for each sample is equal to:

$$a_j = \frac{P_1 - P_0}{P_o} \qquad (1)$$

where $P_1$ is one quarter ahead equity price and $P_0$ is the current equity price.

In total 25 indicators were used to train the regression model. The rank-regression model used in the present study has the following form:

$$a_j = \sum_{i=1}^{25} p_i \times f_{i,j} + e_j \quad (2)$$

where $p_i$ is the coefficient of the feature value $f_i$, $a_j$ is the quarterly-return value, and $e_j$ is the error for the $j^{th}$ sample.

Generally, in equation (2), the sum of squared errors i.e., $\sum_j e_j^2$ is minimised to find the values of coefficients '$p_i$'. Further, if L2 regularisation is being done then the loss takes the following form.

$$Loss = e_j^2 + \sum_{i=1}^{25} p_i^2 \quad (3)$$

Then, this loss function is minimized over all companies to obtain coefficients '$p_i$' values.

However, in the present methodology the following loss-function is minimized to obtain the values of coefficients '$p_i$'.

$$Loss = \sum_{l \in L} \sum_{j \in B_l} (e_j - \frac{\sum_{k \in B_l} e_k}{|B_l|})^2$$

where l denotes the quarter number, $B_l$ denotes the set containing samples belonging to $l_{th}$ quarter, $e_j$ is as described in equation (2) and $|B_l|$ is the cardinality of the set $B_l$.

Here, mean of error of each batch is subtracted from each company's error value ($e_j$). Thus, due to some unpredictable events, a particular quarter's performance is very poor or very good, then this mean-subtraction term will neutralize that effect from the regression equation.

In other words, if the target value in all companies of any 1 quarter are all changed by a constant, still there would be no change in the parameters value. The parameters need not learn/document that change, and get more flexibility/freedom to capture the desired comparative-variation.

Further, if L2 regularisation is being done then the loss takes the following form:

$$Loss = \sum_{l \in L} \sum_{j \in B_l} (e_j - \frac{\sum_{k \in B_l} e_k}{|B_l|})^2 + \sum_{i=1}^{25} p_i^2$$

Then, this loss function is minimized over all companies/samples to obtain coefficients '$p_i$' values.

The missing values in features of the dataset are replaced with zero. The missing values totalled up to 5.2% of all values in the NSE data set. The feature 'Hist To tDebt Comm Eqty Pct' (Historical Total Debt to Common Equity percentage) alone has missing values totalling up to 2%. Four features ('Hist To tDebt Comm Eqty Pct', 'PE' (Price per Earnings), 'Dividend Yield', 'Price To CF Per Share') accounted for 80% of all missing values. In the BSE dataset, which is much smaller than NSE data-set, missing values totalled up to 13.2% of all dataset values. The maximum missing values are in the feature 'Hist To tDebt Comm Eqty Pct' totalling up to 2.7% of all dataset values. The four features, corresponding to the economy, namely 'USD/INR' (US Dollar/ Ruppee Spot rate ), 'IN10YT=RR' (Indian government 10 year bond yield), 'INRPM=RBIA' (India Repo Rate Liquidity Adjustment Facility) and 'MCGBc1' (Indian Crude Oil Energy Future) have no missing values in any of the samples in both the data sets.

Then features are normalized, i.e., the mean and standard deviation of features in training dataset are made equal to 0 and 1 respectively.

### 3.1 Parameters Training Procedure

The parameters are not optimised to predict the actual return. Instead, the parameters are trained in such a way that model should rank the companies by expected return for each quarter. This is done in the following way. Samples in the training data are divided based on the quarterly time period. Thus, samples containing every company's information for a given quarter are all kept in a single batch. Thus, each quarter is represented by a batch which contains relevant information regarding every company for that quarter. Forecasts are generated for each batch and compared with the target/actual values to find the loss. Then, the forecasts and the target values are centralized to zero. The final loss is calculated. Then, the gradients are calculated for the loss with respect to each parameter. The training-procedure followed here, updates the parameters set–wise, so that the complete loss of any one set is reduced in each epoch. The parameters are updated based on their gradients to minimize the final error.

**PSEUDO CODE**

*for each batch/quarter (till the stopping criteria):*

    *N = number of companies in this batch*

    *for each company in this batch do:*

        *$o_i$ = regression output of model for company i*

        *$a_i$ = $i^{th}$ company's quarterly return*

    *for each company in this batch do:*

        *Centralized output ($O_i$) = $o_i - (\sum_{i \in companies} o_i)/N$*

        *Centralized return value ($A_i$) = $a_i - (\sum_{i \in companies} a_i)/N$*

    *Error = $\sum_i (O_i - A_i)^2$*

    *perform Gradient Descent over the model parameters to minimize the final error*

Figure 1: Proposed Algorithm, Pseudo code for the proposed algorithm used to train the model parameters through Gradient Descent (GD).

Figure 1 gives the pseudo-code for the proposed algorithm. Let us call the step containing the term 'centralised' in the above pseudo-code as batch centralisation. This term 'batch centralisation' has been inspired from the work [14] from deep neural networks literature. In [14], batch normalization is done in between deep neural network layers which significantly improves the performance of the model. In present work, batch centralization is done to tide over things which affect the whole stock market. Such things like foreign influence, major revisions in government policies, weather etc affect the whole market and their influence needs to be neutralised. Batch centralization is much more than simply centralizing the target values. Here the parameters have flexibility that they may shift the prediction on any batch/quarter by any constant throughout, still it does not increases the error. Thus, training is done batch-wise with centralisation as explained earlier. The model ranks the companies based on the expected return for the desired quarter. While testing the model, the output values are used to rank the companies for each quarter. The company with a higher output value is ranked at the corresponding higher place. Finally, top companies based on this ranking are chosen for the portfolio.

**3.2 Experimentation Details**

Jiahan et al. [6] compared the performance of regression models with different regularisations (L1 regularisation, L2 regularisation or both). They reported that Ridge KS regression, which uses L2 regularisation, performs best amongst these 3 models during the period of expansion. Expansion denotes the period of the economy where stock prices generally grow higher, as defined by the National Bureau of Economic Research (NBER), USA. The data set used in the present study is of the period when stocks increased steadily and the Indian economy expanded.

Owing to the reported success of regularisation in such task [6], L2 regularisation is done to the trainable parameters while being trained through Stochastic gradient descent. This means that the sum of squares of the coefficients is added to the final loss function. Thus, the coefficients are trained with the dual objective of obtaining correct forecasts and keeping the coefficient size small. The results obtained through the proposed rank-regression model with regularisation are reported separately from the results of the rank-regression model. Finally, the proposed models are compared with a simple regression model, a ridge regression model and a naïve model. The simple regression model and ridge regression model used the same input features to train a linear relationship between the inputs and the quarterly returns, as in the proposed models. These models differ only in the way the parameters are learned on the training set. The simple regression and ridge regression model used Stochastic Gradient Descent (SGD) method to train the parameters of the equation. The SGD method is suitable for big datasets and it is even extendable to very large datasets. Further, using the same SGD method for all models, validated the advantage of batch centralisation technique used in the proposed model. The naïve model ranked the companies based on the previous quarter returns. The best-performing companies in the previous quarter are expected to perform best in the current quarter as well. This is inspired by the Wiener process or random-walk based modelling of stock prices. The experiments were performed on these five models to ascertain the best models amongst these. Table 1 describes the models, leaving out the naïve model.

In each experiment, a model is trained on the past quarter's information and is used to rank all the listed stocks in the next quarter. An expected relative return is generated by the proposed model for each stock used for training. This relative return finally determines the rank or performance of that stock amongst all others. Here, 25 parameters need to be learnt for each of the dataset consisting of either 497 companies or 708 companies. This means that these parameters should minimise the error in all of these 497 or 708 companies simultaneously. This has to be done several times, during cross-validation experiments as explained later. Therefore, the dataset is large and requires huge computation for calculating parameters. Thus, stochastic gradient descent methodology is used for parameters training. This is the same methodology which is used for learning deep neural networks, where it is more sophistically referred to as back propagation algorithm.

The error is calculated for each stock, and the parameters are updated to minimise that error. This is done until the stopping criteria is reached. The stopping criteria stop the iteration procedure whenever the error has reached its minimum value given one condition. The condition is that before stopping, the current minimum error value has not reduced further for the last thirty thousand iterations. The maximum number of iterations is capped at eight lakhs though iterations stopped mostly before that number.

The algorithms were coded into python 3 for experimentation. Pytorch, a python deep learning module, was utilised for learning the regression parameters.

Next, we provide information regarding the datasets, validation strategy and evaluation metrics.

### 3.2.1 Dataset description

Two datasets are extracted from Thomson Reuters Eikon tool to perform the experiments. The first data set consists of 497 companies listed at BSE 500 index. The second data set consists of 708 companies traded at National Stock Exchange (NSE) of India. The companies' information are extracted for a period of seven years for each quarter. The period considered starts from the January-March quarter of 2011 and ends at October-December quarter of 2017. Thus in total, the data sets contains information and a target value for companies for these 28 quarters.

### 3.2.2 Evaluation metrics

The models mentioned above are used to rank the stocks by expected next quarter return. The models are compared on the basis of five metrics:

1.) **AP@100** : Average Precision Metric for top 100 recommendations (AP@100). This measure is given by:

$$AP@100 = \frac{1}{100} \times \sum_{k=1}^{100} P(k).rel(k) \quad (3)$$

where P(k) is obtained after dividing 'number of correct recommendations amongst top k companies' by 'k'. 'rel(k)' is an indicator function which is 1 when $k^{th}$ recommendation is amongst top k and otherwise 0. This measure has been chosen as the main concern is to be precise about recommending top companies in the correct order.

**Table 1**: Model descriptions of different models. Feature normalization refers to normalization of input features. Target value normalization refers to normalization of target value using mean and standard deviation of the training data target values. Thus, training data target values mean and standard deviation becomes equal to 0 and 1 respectively. But, test data target values mean and standard deviation is not exactly equal to 0 and 1 respectively. '$\bar{o}$' and '$\bar{a}$' refers to mean of model output values and actual values for a batch respectively. '$p_i$' refers to regression equation coefficient value.

|   | Model | Pre-processing steps | Final error function | Training Procedure |
|---|---|---|---|---|
| 1 | Simple SGD | Feature Normalisation + Target value Normalisation | $\sum(o_i - a_i)^2$ | Parameters are updated through SGD after each iteration consisting of only 1 sample. |
| 2 | Simple SGD + L2 | Feature Normalisation + Target value Normalisation | $\sum(o_i - a_i)^2 + \frac{1}{2}\sum p_i^2$ | Parameters are updated through SGD after each iteration consisting of only 1 sample. |
| 3 | Rank-regression | Feature Normalisation | $\sum((o_i - \bar{o}) - (a_i - \bar{a}))^2$ | Parameters are updated through SGD after each iteration consisting of all samples within each quarter. |
| 4 | Rank-regression + L2 | Feature Normalisation | $\sum((o_i - \bar{o}) - (a_i - \bar{a}))^2 + \frac{1}{2}\sum p_i^2$ | Parameters are updated through SGD after each iteration consisting of all samples within each quarter. |

2.) **Top 20** : Actual Return generated by investing equally in top 20 stocks as recommended by a model
3.) **Top 50** : Actual Return generated by investing equally in top 50 stocks as recommended by a model
4.) **Riskless 20** : Non-dominated solutions (optimal solutions) with low risk and high return are calculated. Riskless 20 gives mean-return of top 20 stocks lying on the front of least risk and highest return. The standard deviation of past returns was used as a risk measure. Within each front, stocks were sorted according to the higher return value.
5.) **Riskless 50** : Similar to Riskless 20, Riskless 50 gives mean-return of top 50 stocks lying on the front of least risk and highest return.

### 3.2.3 Validation Strategy

Experiments are conducted on different models through extended cross-validation. The cross-validation is done through organising data into rolling windows, where training is done on a certain number of consecutive quarters and the next quarter is used for testing purpose. The number of consecutive quarters used for training in 3 different experiments are 10, 15 and 20. Thus in total results are validated on 18, 13 and 8 pairs of train-test data sets corresponding to experiments with 10, 15 and 20 training-timestamps/training-quarters respectively. Here, 18 refers to 18 windows obtained by rolling the window from the first eleven quarters to the last eleven quarters of the total 28 quarters. Likewise, other numbers 13 and 8 can be deciphered. Tables 2-7 give the average results for each of these cross-validation sets.

## 4. Results

The overall results of the experiments demonstrate the superiority of the proposed models. Table 2,3 show results when 10 quarters used for learning the parameters. Similarly, Table 4,5 correspond to 15 quarters and table 6,7 correspond to 20 quarters.

**Table 2:** Results corresponding to BSE Data set and 10 number of training quarters.

|  | Evaluation Metrics | | | | |
|---|---|---|---|---|---|
| **Model Name** | AP@100 | Top 20 | Top 50 | Riskless 20 | Riskless 50 |
| Rank-Regression (RR) + L2 | **0.03256** | **0.16751** | **0.14684** | 0.11672 | **0.11245** |
| RR | 0.02736 | 0.12361 | 0.12747 | 0.10809 | 0.09784 |
| Ridge | 0.03172 | 0.16544 | 0.14177 | **0.12192** | 0.10888 |
| Simple SGD | 0.02914 | 0.14011 | 0.13848 | 0.10535 | 0.10194 |
| Naïve | 0.01303 | 0.08533 | 0.07882 | 0.06308 | 0.06490 |

**Table 3:** Results corresponding to NSE Data set and 10 number of training quarters.

|            | Evaluation Metrics |         |         |             |             |
|------------|--------------------|---------|---------|-------------|-------------|
| Model Name | AP@100             | Top 20  | Top 50  | Riskless 20 | Riskless 50 |
| RR + L2    | **0.01590**        | 0.13808 | 0.13250 | **0.14037** | **0.11722** |
| RR         | 0.01492            | 0.13885 | 0.12970 | 0.11910     | 0.09885     |
| Ridge      | 0.01547            | **0.13974** | 0.13435 | 0.14015 | 0.11270     |
| Simple SGD | 0.01522            | 0.13606 | **0.13686** | 0.13632 | 0.10873   |
| Naïve      | 0.00513            | 0.08994 | 0.07334 | 0.05761     | 0.07052     |

**Table 4:** Results corresponding to BSE Data set and 15 number of training quarters.

|            | Evaluation Metrics |         |         |             |             |
|------------|--------------------|---------|---------|-------------|-------------|
| Model Name | AP@100             | Top 20  | Top 50  | Riskless 20 | Riskless 50 |
| RR + L2    | **0.03203**        | 0.11631 | **0.10228** | 0.07355 | **0.08015** |
| RR         | 0.03183            | 0.11644 | 0.09927 | **0.08656** | 0.07329     |
| Ridge      | 0.03191            | **0.12486** | 0.09808 | 0.07647 | 0.07915     |
| Simple SGD | 0.02888            | 0.11669 | 0.09660 | 0.07924     | 0.07772     |
| Naïve      | 0.01300            | 0.04686 | 0.04191 | 0.04135     | 0.03756     |

**Table 5:** Results corresponding to NSE Data set and 15 number of training quarters.

|            | Evaluation Metrics |         |         |             |             |
|------------|--------------------|---------|---------|-------------|-------------|
| Model Name | AP@100             | Top 20  | Top 50  | Riskless 20 | Riskless 50 |
| RR + L2    | 0.01643            | 0.09844 | 0.10033 | 0.09247     | 0.07627     |
| RR         | **0.01882**        | **0.11941** | **0.10646** | **0.09991** | **0.08249** |
| Ridge      | 0.01665            | 0.09839 | 0.09586 | 0.09363     | 0.07869     |
| Simple SGD | 0.01704            | 0.10528 | 0.10235 | 0.07885     | 0.05952     |
| Naïve      | 0.00689            | 0.06343 | 0.05698 | 0.02914     | 0.04165     |

**Table 6:** Results corresponding to BSE Data set and 20 number of training quarters.

|            | Evaluation Metrics |         |         |             |             |
|------------|--------------------|---------|---------|-------------|-------------|
| Model Name | AP@100             | Top 20  | Top 50  | Riskless 20 | Riskless 50 |
| RR + L2    | **0.03276**        | **0.15115** | **0.12630** | 0.10537 | **0.10268** |
| RR         | 0.03219            | 0.13716 | 0.12328 | 0.10072     | 0.09683     |
| Ridge      | 0.03134            | 0.15018 | 0.12572 | **0.10661** | 0.10162     |
| Simple SGD | 0.02720            | 0.12734 | 0.12326 | 0.10460     | 0.09768     |
| Naïve      | 0.01107            | 0.08106 | 0.06676 | 0.06829     | 0.05948     |

**Table 7:** Results corresponding to NSE Data set and 20 number of training quarters.

|            | Evaluation Metrics |         |         |             |             |
|------------|--------------------|---------|---------|-------------|-------------|
| Model Name | AP@100             | Top 20  | Top 50  | Riskless 20 | Riskless 50 |
| RR + L2    | 0.01753            | 0.11727 | 0.12319 | 0.10021     | 0.10575     |
| RR         | **0.02116**        | **0.12968** | 0.11513 | 0.08292 | 0.08120     |
| Ridge      | 0.01782            | 0.12102 | **0.12389** | **0.11436** | **0.10833** |
| Simple SGD | 0.02099            | 0.12514 | 0.11699 | 0.09418     | 0.08668     |
| Naïve      | 0.00650            | 0.06569 | 0.06514 | 0.04615     | 0.05956     |

The metric AP@100 is consistently higher in all the three experiments for the proposed models. The results for NSE data set on 15 quarters particularly stand out because here the rank-regression model without regularisation outperforms all the rest models in all the evaluation metrics.

In Tables 2-7, the proposed models outperform the existing models in at least three of the evaluation metrics except Table 7. In Table 7 which corresponds to NSE data set trained on 20 quarters, still, AP@100 and Top 20 returns metrics point towards the superiority of the rank-regression models.

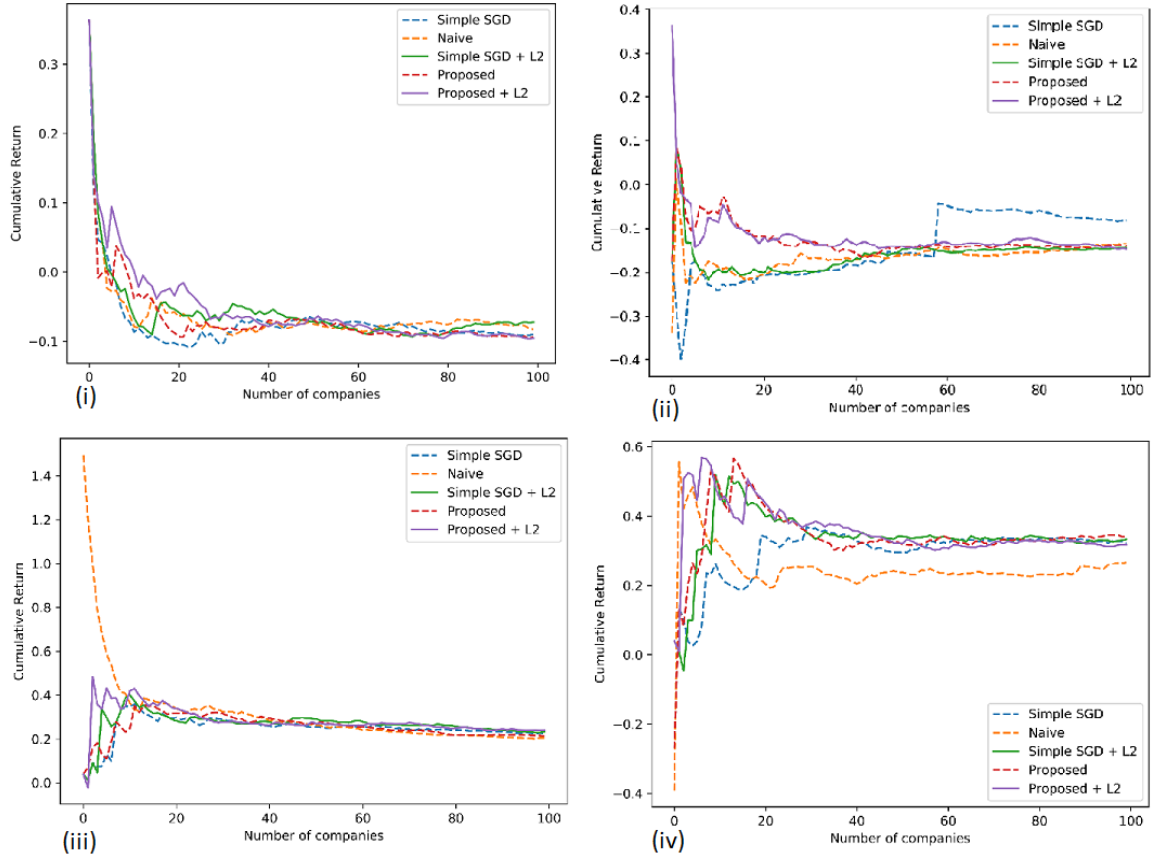

**Figure 2**: Cumulative Mean Returns, Mean return of companies ranked according to expected return as per each of the models. The number of companies plotted for each model has been truncated at 100. Each sub-figure denotes different dataset or period, i.e., (i) Q3 2017-18, BSE dataset, (ii) Q4 2017-18, NSE dataset, (iii) Q4 2017-18, BSE dataset, (i) Q3 2017-18, NSE dataset.

Further analysis is done on results obtained by training the model on 26 quarters and testing on the last two quarters. Here the results are subjectively analysed through figures. The figures elaborate the significance of the proposed models. The numerical results corresponding to this experiment are given in Table 1A and 2A (Appendix)

Figures 2 and 1A (Appendix) give the plot of cumulative mean-return for companies ranked according to higher output in each model. The steps to obtain each plot are:

1. For each model, we obtain the ranks for each company by sorting as per the expected relative return generated by the model.
2. Then we take the mean of actual returns for top 'n' companies.
3. This gives us the y-axis value corresponding to value 'n' on x-axis.
4. The y-axis values correspond to the returns generated when an equal amount of money would have been invested in each of the top 'n' companies.

In figures 2 and 1A (Appendix), the proposed models' cumulative-mean returns are mostly higher than that of other models. This means that equal investment in top companies recommended by the proposed models generates more profit than investing in companies recommended by other models. The cumulative mean return remains high for the initial few companies and then gently slopes down to the lowest value at the end. This curve perfectly depicts that the model has been able to rank the companies correctly. Companies having higher rank do tend to show a higher return in that quarter. After 100 companies, mean cumulative return becomes more or less the same for all the models. Thus, the proposed model has been successfully able to single out top performers amongst different stocks.

## 5. Disussion

The present study uses a simple portfolio investment strategy of investing equally in all the top recommended stocks. Mean-variance trading strategy has been used in past researches [6] to create a portfolio based on expected return and variance of each stock. But the present study cannot use that strategy as expected return forecasts are not generated by the proposed model. It recommends top companies and then invests equally in the top few companies.

Stochastic gradient descent (Back propagation) method with batch update has been preferred so that parameters are updated in accordance with variability in each batch/quarter. This is because the gradient of loss of the whole batch is used to update parameters repeatedly at each update. Still, comparisons have been obtained, which showcase rank regression models effectiveness.

Ensemble models are predicting models which combine forecasts from two or more models to get final prediction value. They generally perform better than individual models and there exist theoretical and empirical foundations [15] for this result. The studies [16]–[18] used hybrid models to achieve better performance for prediction. In an extension of this work, the proposed model may be combined with other techniques to achieve superior results.

## 6. Conclusion

The present paper proposes a novel technique for recommending top stocks in any period. The centralization technique employed in the training procedure has not been used earlier in equity-forecasting literature. Also, comparing stocks in each quarter separately leads to better stock predictions. In contrast, generally, researchers simply fit a regression equation on the equity features, which is not a very robust technique. Two large data sets consisting of 497 companies listed at BSE (India) and 708 companies listed at NSE (India), are used for the experiments. The results were evaluated using five evaluation metrics. The results proved the superiority of the proposed technique. Results are also subjectively analysed through figures, which clearly show the effectiveness of the rank-regression. The research paper also verified that L2 regularisation is useful in improving equity forecast results for the Indian stock market, as reported in past studies for other markets.

## Acknowledgements

This work was not supported by any agency.

## References

[1]     D. E. Rapach, J. K. Strauss, and G. Zhou, "Out-of-Sample Equity Premium Prediction: Combination Forecasts and Links to the Real Economy," *Rev. Financ. Stud.*, vol. 23, no. 2, pp. 821–862, Feb. 2010.
[2]     I. Welch and A. Goyal, "A Comprehensive Look at The Empirical Performance of Equity Premium Prediction," *Rev. Financ. Stud.*, vol. 21, no. 4, pp. 1455–1508, Jul. 2008.
[3]     J. Y. Campbell and S. B. Thompson, "Predicting Excess Stock Returns Out of Sample: Can Anything Beat the Historical Average?," *Rev. Financ. Stud.*, vol. 21, no. 4, pp. 1509–1531, Jul. 2008.
[4]     T. Dangl and M. Halling, "Predictive regressions with time-varying coefficients," *J. financ. econ.*, vol. 106, no. 1, pp. 157–181, Oct. 2012.
[5]     C. J. Neely, D. E. Rapach, J. Tu, and G. Zhou, "Forecasting the Equity Risk Premium: The Role of Technical Indicators," *Manage. Sci.*, vol. 60, no. 7, pp. 1772–1791, Jul. 2014.
[6]     J. Li, I. Tsiakas, J. Li, and I. Tsiakas, "Equity Premium Prediction: The Role of Economic and Statistical Constraints," *J. Financ. Mark.*, vol. 36, pp. 56–75, 2017.
[7]     S. Giebel and M. Rainer, "Neural network calibrated stochastic processes: forecasting financial assets," *Cent. Eur. J. Oper. Res.*, vol. 21, no. 2, pp. 277–293, Mar. 2013.
[8]     G. Elliott and A. Gargano, "Complete subset regressions," *J. Econom.*, vol. 177, no. 2, pp. 357–373, Dec. 2013.
[9]     R. Bhar and B. Nikolova, "Oil Prices and Equity Returns in the BRIC Countries," *World Econ.*, vol. 32, no. 7, pp. 1036–1054, Jul. 2009.
[10]    V. Tripathi, "Company Fundamentals and Equity Returns in India," *SSRN Electron. J.*, May 2008.
[11]    L. C. Bhandari, "Debt/Equity Ratio and Expected Common Stock Returns: Empirical Evidence," *J. Finance*, vol. 43, no. 2, pp. 507–528, Jun. 1988.
[12]    T. Dangl and M. Halling, "Predictive regressions with time-varying coefficients," *J. financ. econ.*, vol. 106, no. 1, pp. 157–181, 2012.
[13]    J. Li, I. Tsiakas, and W. Wang, "Predicting Exchange Rates Out of Sample: Can Economic Fundamentals Beat the Random Walk?," *e J. Financ. Econom.*, 2014.
[14]    S. Ioffe and C. Szegedy, "Batch Normalization: Accelerating Deep Network Training by Reducing Internal Covariate Shift," 2015.
[15]    R. Hagedorn, F. J. Doblas-Reyes, and T. N. Palmer, "The rationale behind the success of multi-model

ensembles in seasonal forecasting — I. Basic concept," *Tellus A Dyn. Meteorol. Oceanogr.*, vol. 57, no. 3, pp. 219–233, Jan. 2005.
[16] G. P. Zhang, "Time series forecasting using a hybrid ARIMA and neural network model," *Neurocomputing*, vol. 50, pp. 159–175, 2003.
[17] P. R. A. Firmino, P. S. G. de Mattos Neto, and T. A. E. Ferreira, "Error modeling approach to improve time series forecasters," *Neurocomputing*, vol. 153, pp. 242–254, 2015.
[18] M. Khashei and M. Bijari, "A novel hybridization of artificial neural networks and ARIMA models for time series forecasting," *Appl. Soft Comput.*, vol. 11, no. 2, pp. 2664–2675, 2011.